\documentclass{PoS}

\usepackage{graphicx}% Include figure files
\usepackage{dcolumn}% Align table columns on decimal point
\usepackage{bm}% bold math
\usepackage{cite}
\usepackage{amsmath, amssymb}% for complicated equations

\newcommand{\Psfig}[2]{\includegraphics[width=#1]{#2}}

\def\prt{\partial}

\def\mev{\text{ MeV}}

\def\fm{\text{ fm}}

\title{Two-body Wave Functions, Compositeness, And The Internal
  Structure Of Dynamically Generated Resonances}

\ShortTitle{Two-body Wave Functions, Compositeness, And The Internal
  Structure}% Of Dynamically Generated Resonances}

\author{\speaker{Takayasu Sekihara}%\thanks{A footnote may follow.}
  \\
  Advanced Science Research Center, Japan Atomic Energy Agency, 
  Shirakata, Tokai, Ibaraki, 319-1195, Japan\\
  E-mail: \email{sekihara@post.j-parc.jp}}

\author{Tetsuo Hyodo\\
  Yukawa Institute for Theoretical Physics, Kyoto
  University, Kyoto 606-8502, Japan\\
  E-mail: \email{hyodo@yukawa.kyoto-u.ac.jp}}

\author{Daisuke Jido\\
  Department of Physics, Tokyo Metropolitan University,
  Hachioji 192-0397, Japan\\
  E-mail: \email{jido@tmu.ac.jp}}

\author{Junko Yamagata-Sekihara\\
  National Institute of Technology, Oshima College, Oshima,
  Yamaguchi, 742-2193, Japan\\
  E-mail: \email{yamagata@oshima-k.ac.jp}}

\author{Shigehiro Yasui\\
  Department of Physics, Tokyo Institute of Technology,
  Tokyo 152-8551, Japan\\
  E-mail: \email{yasuis@th.phys.titech.ac.jp}}

\abstract{Recently, the compositeness, defined as the norm of a
  two-body wave function for bound and resonance states, has been
  investigated to discuss the internal structure of hadrons in terms
  of hadronic molecular components.  From the studies of the
  compositeness, it has been clarified that the two-body wave function
  of a bound state can be extracted from the residue of the scattering
  amplitude at the bound state pole.  Of special interest is that the
  two-body wave function from the scattering amplitude is
  automatically normalized.  In particular, while the compositeness is
  unity for energy-independent interactions, it deviates from unity
  for energy-dependent interactions, which can be interpreted as a
  missing-channel contribution.  In this manuscript, we show the
  formulation of the two-body wave function from the scattering
  amplitude, evaluate the compositeness for several dynamically
  generated resonances such as $f_{0} (980)$, $\Lambda (1405)$, and
  $\Xi (1690)$, and investigate their internal structure in terms of
  the hadronic molecular components.}

\FullConference{The 26th International Nuclear Physics Conference\\
		11-16 September, 2016\\
		Adelaide, Australia}

\begin{document}

\section{Introduction}

Thanks to the recent improvements of the hadron
spectroscopy~\cite{Olive:2016xmw}, we can discuss not only global
quantities of hadrons such as masses and widths but also their
internal structure.  In particular, from experimental observables we
can examine various hadrons for their exotic configurations such as
multi-quark states, dibaryons, and hadronic molecules instead of the
ordinary configurations, {\it i.e.}, three quarks for baryons and a
quark--antiquark pair for mesons.

In this line, hadronic molecular configuration is of special interest
because the structure of hadrons can be expressed in terms of the
hadronic degrees of freedom, which are asymptotic states of the
fundamental theory of strong interactions, QCD, as distinguishable
components~\cite{Hyodo:2013nka}.  Actually, from decades ago,
hadron--hadron scattering amplitudes have been utilized for clarifying
the structure of states in terms of hadronic
molecules~\cite{Weinberg:1965zz, Baru:2003qq}.  Recently,
hadron--hadron scattering amplitudes have been intensively studied so
as to investigate the hadronic molecular components for hadronic
resonances in the amplitude, and these studies can be characterized by
the compositeness~\cite{Hyodo:2013nka, Hyodo:2011qc, Aceti:2012dd,
  Hyodo:2013iga, Nagahiro:2014mba, Aceti:2014ala, Sekihara:2014kya,
  Kamiya:2015aea, Sekihara:2015gvw, Guo:2015daa}, defined as the norm
of a two-body wave function interest, based on the fact that the wave
function extracted from the scattering amplitude is automatically
normalized~\cite{Hernandez:1984zzb, Gamermann:2009uq,
  Sekihara:2016xnq}.

In the present manuscript, we formulate the two-body wave functions
and compositeness from the scattering amplitudes and discuss the
internal structure of several dynamically generated hadronic
resonances in terms of the hadronic molecular
components~\cite{Sekihara:2014kya, Sekihara:2015gvw,
  Sekihara:2016xnq}.

\section{Two-body wave functions and compositeness from scattering
  amplitudes}

First of all, we consider a quantum system governed by the Hamiltonian
$\hat{H} = \hat{H}_{0} + \hat{V}$ with the free part $\hat{H}_{0}$ and
interaction $\hat{V}$.  We assume that the present model space is
restricted to two-body states.  The free Hamiltonian $\hat{H}_{0}$ has
a two-body eigenstate of its relative momentum $\bm{q}$ as:
\begin{equation}
  \hat{H}_{0} | \bm{q}_{j} \rangle = \mathcal{E}_{j} ( q ) | \bm{q}_{j} \rangle
  , 
  \quad 
  \mathcal{E}_{j} ( q ) \equiv 
%  \begin{cases}
%    \displaystyle m_{j} + M_{j} + \frac{q^{2}}{2 \mu _{j}} 
%    & \text{(Non-relativistic case)} ,
%    \\
%    \displaystyle \sqrt{q^{2} + m_{j}^{2}} + \sqrt{q^{2} + M_{j}^{2}}
%    & \text{(Semi-relativistic case)} ,
%  \end{cases}
  \sqrt{q^{2} + m_{j}^{2}} + \sqrt{q^{2} + M_{j}^{2}}
  \quad \text{(Semi-relativistic case)} ,
\end{equation}
where $q \equiv | \bm{q} |$, $j$ is the channel index, and $m_{j}$ and
$M_{j}$ are masses of the particles in channel $j$.  With this
eigenstate, the interaction can be evaluated as $V_{j k} ( E ; \,
\bm{q}^{\prime} , \, \bm{q} ) = \langle \bm{q}_{j}^{\prime} | \hat{V}
( E ) | \bm{q}_{k} \rangle$, where the interaction is allowed to
depend intrinsically on the energy of the system $E$.

In this construction, we can consider a two-body to two-body
scattering process $k ( \bm{q} ) \to j ( \bm{q}^{\prime} )$.  The
scattering amplitude for this process, $T_{j k} ( E ; \,
\bm{q}^{\prime} , \, \bm{q} )$, is a solution of the
Lippmann--Schwinger equation, and after the projection to the partial
wave of orbital angular momentum $L$ we have
\begin{equation}
  T_{L, \, j k} ( E ; \, q^{\prime} , \, q ) 
  = V_{L, \, j k} ( E ; \, q^{\prime} , \, q ) 
  + \sum _{l} \int \frac{d k}{2 \pi ^{2}} k^{2}
  \frac{V_{L, \, j l} ( E ; \, q^{\prime} , \, k ) 
    T_{L, \, l k} ( E ; \, k , \, q  )}{E - \mathcal{E}_{l} ( k )} .
  \label{eq:LSint}
\end{equation}

If the interaction generates a bound state, including an unstable
case, the scattering amplitude has a corresponding pole in the complex
energy plane.  In particular, for an off-shell
amplitude, which we may treat as a function of three independent
variables $E$, $q^{\prime}$, and $q$, the pole can be expressed as
\begin{equation}
  T_{L, \, j k} ( E ; \, q^{\prime} , \, q ) = \frac{\gamma _{j} ( q^{\prime} ) 
    \gamma _{k} ( q )}{E - E_{\rm pole}} + (\text{regular at } E =E_{\rm pole} ) ,
\end{equation}
with the residue with respect to the pole at $E = E_{\rm pole}$,
$\gamma _{j} ( q )$.

Interestingly, the residue $\gamma _{j} ( q )$ contains information on
the wave function of the bound state $| \Psi \rangle$.  Actually, in
the vicinity of the pole position, we can evaluate the scattering
amplitude in the expansion by the eigenstates of the full Hamiltonian,
which results in
\begin{equation}
  T_{L, j k} ( E ; \, q^{\prime} , \, q ) \approx
  \langle \bm{q}_{j}^{\prime} | \hat{V} ( E_{\rm pole} ) | \Psi \rangle
  \frac{1}{E - E_{\rm pole}} \langle \tilde{\Psi} | \hat{V} ( E_{\rm pole} )
  | \bm{q}_{k} \rangle ,
\end{equation}
where we have introduced a Gamow vector $\langle \tilde{\Psi} |$ so as
to establish the normalization of the resonance
states~\cite{Gamow:1928zz, Hyodo:2013nka, Hernandez:1984zzb}.  Here
$\langle \tilde{\Psi} | \Psi \rangle = 1$ is guaranteed by taking the
residue of the propagator $( E - E_{\rm pole} )^{-1}$, which is
nothing but the field renormalization constant for the bound state,
unity.  Then, the residue $\gamma _{j} ( q )$ is calculated as
\begin{equation}
  \gamma _{j} ( q ) 
  = \langle \bm{q}_{j} | \hat{V} ( E_{\rm pole} )  | \Psi \rangle
  = [ E_{\rm pole} - \mathcal{E}_{j} ( q ) ] \langle \bm{q}_{j} | \Psi \rangle ,
\end{equation}
and similarly for $\langle \tilde{\Psi} | \hat{V} ( E_{\rm pole} ) |
\bm{q}_{j} \rangle$.  An important property is that the residue
$\gamma _{j} ( q )$ as well as the scattering amplitude should be
automatically normalized because the Lippmann--Schwinger equation is
an inhomogeneous integral equation.  This fact is especially essential
when we calculate the norm of the $j$th channel component of the wave
function, $\langle \bm{q}_{j} | \Psi \rangle$:
\begin{equation}
  X_{j} \equiv \int \frac{d^{3} q}{( 2 \pi )^{3}} 
  \langle \tilde{\Psi} | \bm{q}_{j} \rangle
  \langle \bm{q}_{j} | \Psi \rangle 
  = \int _{0}^{\infty} d q \, \text{P}_{j} ( q ) , 
  \quad 
  \text{P}_{j} ( q ) \equiv \frac{q^{2}}{2 \pi ^{2}} \left [ 
    \frac{\gamma _{j} ( q )}{E_{\rm pole} - \mathcal{E}_{j} ( q )} \right ]^{2} ,
\end{equation}
where the norm $X_{j}$ is called compositeness.

\begin{figure}[!t]
  \centering 
  \vspace{-15pt}
  ~ ~ \Psfig{7.0cm}{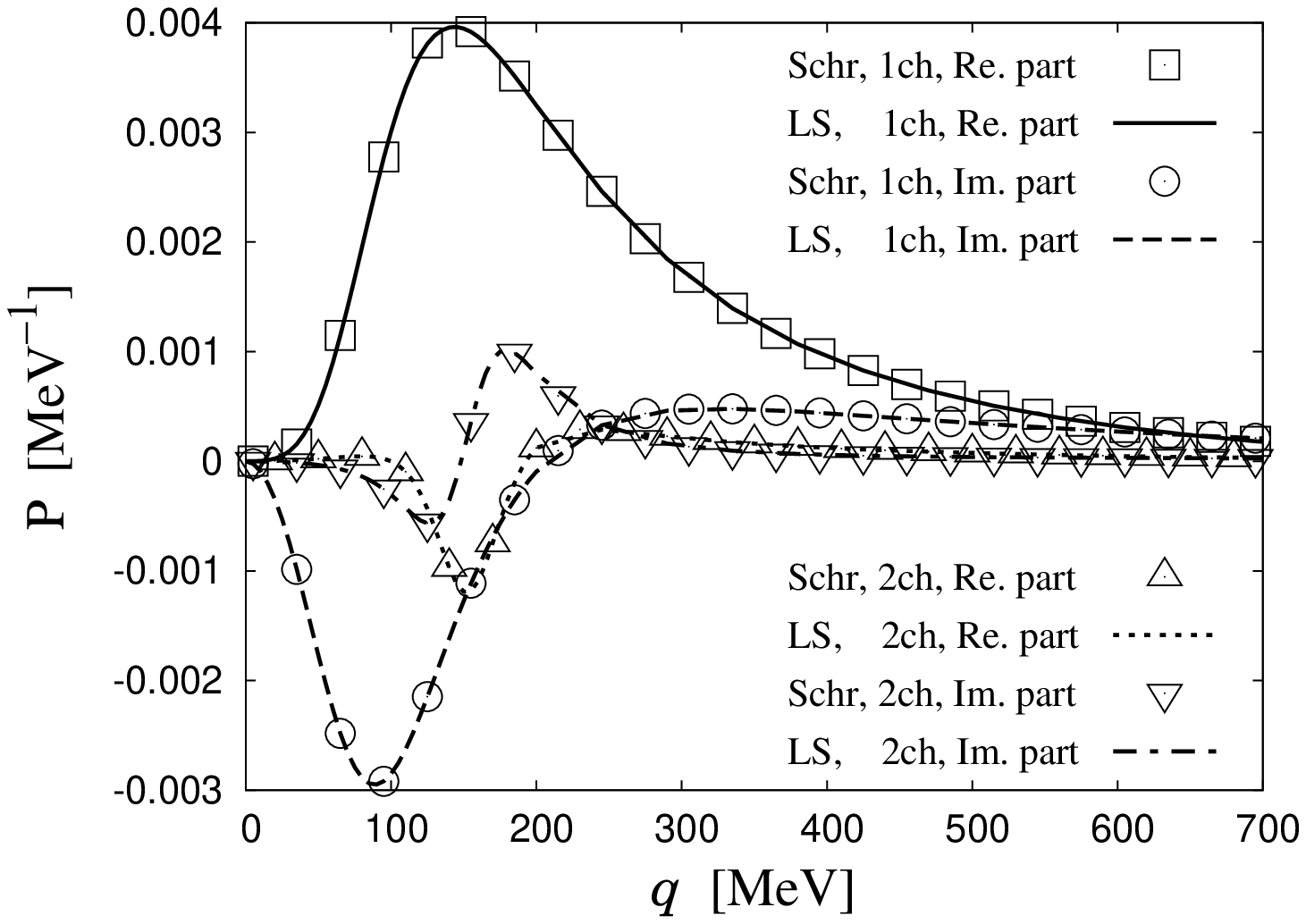} ~ \Psfig{7.0cm}{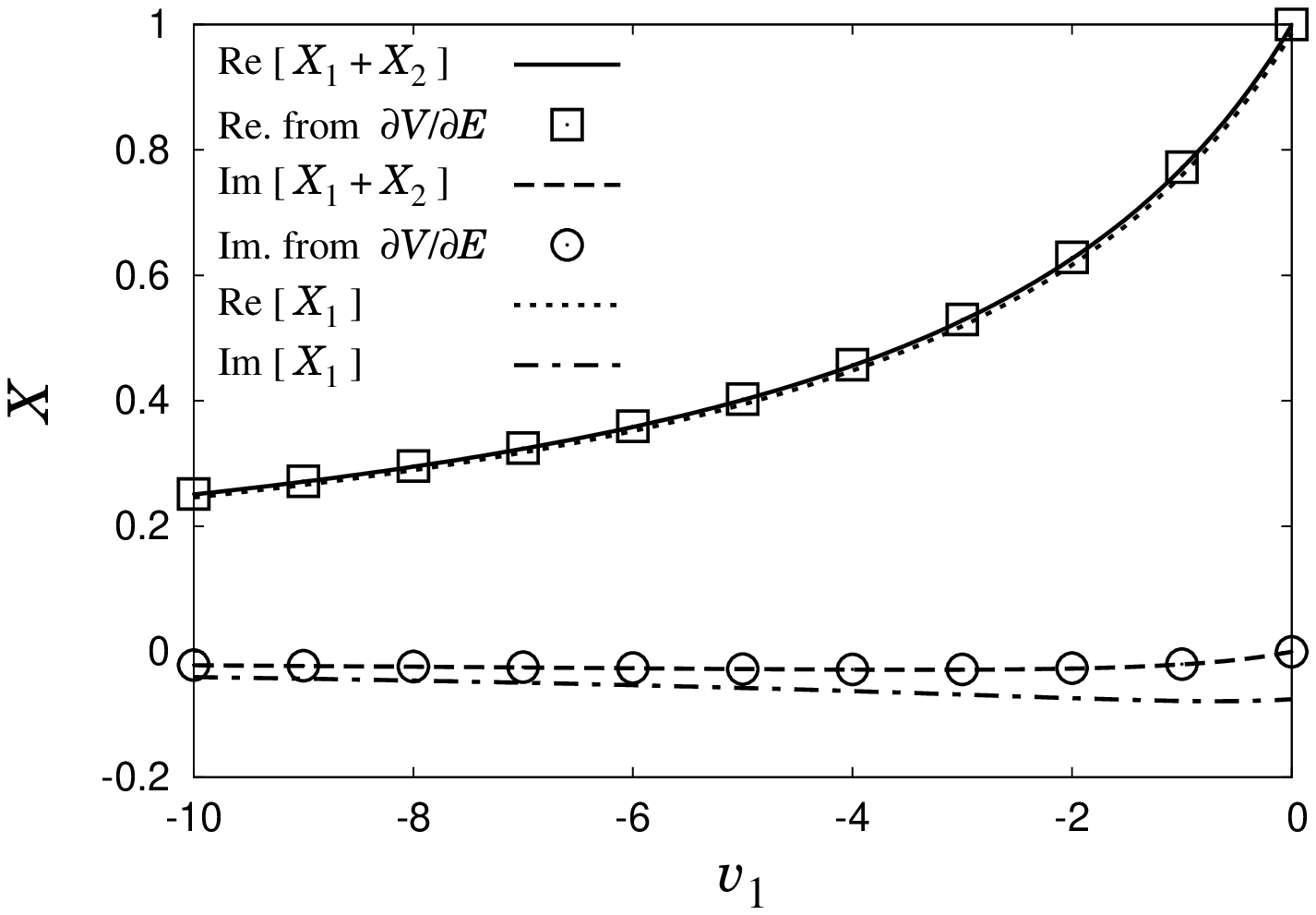}
  \vspace{5pt}
%  \\
%  ~
  \caption{(Left) Density distribution $\text{P}_{j}$ as a function of
    the momentum $q$ with the
    interaction~\eqref{eq:Vjk}~\cite{Sekihara:2016xnq}.  (Right)
    Compositeness as a function of the parameter $v_{1}$ with the
    interaction~\eqref{eq:Vjk}~\cite{Sekihara:2016xnq}.}
  \label{fig:1}
\end{figure}

The correct normalization of $X_{j}$ from the residue of the
scattering amplitude at the pole was proved in
Ref.~\cite{Hernandez:1984zzb} for a general single-channel
energy-independent interaction in the non-relativistic case.  For a
more general case, one can show that the correct normalization of
$X_{j}$ from the amplitude is achieved with some schematic models.
For instance, in Fig.~\ref{fig:1}(left) we show the density
distribution $\text{P}_{j} ( q )$ for a resonance in a two-channel
problem where the interaction is
\begin{equation}
  V_{j k} ( E ; \, \bm{q}^{\prime} , \, \bm{q} )
  = C_{j k} v ( E ) ( \sqrt{\pi} b )^{3} 
  e^{- | \bm{q}^{\prime} - \bm{q} |^{2} b^{2} / 4} ,
  \quad
  C_{j k} = \left ( 
  \begin{array}{@{\,}cc@{\,}}
    1 & x \\
    x & 0 \\
  \end{array}
  \right ) ,
  \quad
  v ( E ) = v_{0} + v_{1} ( E - E_{\rm pole} ) ,
  \label{eq:Vjk}
\end{equation}
with the parameters $b = 0.5 \fm$, $x = 0.5$, $v_{0} = -650 \mev$,
$v_{1} = 0$, and $(m_{1}, \, M_{1} , \, m_{2} , \, M_{2} ) = ( 495.7,$
$938.9 , \, 138.0 , \, 1193.1 ) \mev$.  In Fig.~\ref{fig:1}(left), the
points are obtained by solving the Schr\"{o}dinger equation and
normalizing the wave function ($X_{1} + X_{2} = 1$) by hand in a usual
manner, while the lines by solving the Lippmann--Schwinger equation at
the pole position and extracting the wave function from the scattering
amplitude.  We can see an exact correspondence between the points and
lines for each component in Fig.~\ref{fig:1}(left), which means that
we obtain the correctly normalized wave function from the scattering
amplitude for the energy-independent interaction.

Then, an interesting thing happens when we introduce the energy
dependence of the interaction with $v_{1} \neq 0$.  Actually, with
$v_{1} \neq 0$, the compositeness from the scattering amplitude, which
should be automatically normalized, deviates from unity as seen in the
lines of Fig.~\ref{fig:1}(right), where we plot the compositeness as a
function of $v_{1}$.  One interpretation of this behavior is the
reflection of a missing-channel contribution.  The deviation can be
compared with the compositeness in an energy dependent interaction
discussed in, {\it e.g.}, Refs.~\cite{Formanek:2003,
  Miyahara:2015bya}.  According to these studies, the compositeness
with the energy-dependent interaction is~\cite{Formanek:2003,
  Miyahara:2015bya, Sekihara:2016xnq}
\begin{equation}
  X_{\prt V / \prt E}
  = 1 + \sum _{j , k} \int _{0}^{\infty} d q \frac{q^{2}}{2 \pi ^{2}}
  \frac{\gamma _{j} ( q )}{E_{\rm pole} - \mathcal{E}_{j} ( q )}
  \int _{0}^{\infty} d q^{\prime} \frac{q^{\prime \, 2}}{2 \pi ^{2}}
  \frac{\gamma _{k} ( q^{\prime} )}{E_{\rm pole} - \mathcal{E}_{k} ( q^{\prime} )}
  \frac{\prt V_{L , j k}}{\prt E}
  ( E_{\rm pole} ; \, q, \, q^{\prime} ) ,
  \label{eq:XdVdE}
\end{equation}
so that the continuity equation from the wave function should be hold.
This compositeness $X_{\prt V / \prt E}$ is shown in
Fig.~\ref{fig:1}(right) as points, which lies exactly on the lines of
the real and imaginary parts of $X_{1} + X_{2}$.  This indicates that
the two-body wave function from the scattering amplitude correctly
takes into account the effect of the additional second term
in~\eqref{eq:XdVdE} due to the energy dependence of the interaction.

\section{Compositeness for dynamically generated hadronic resonances}

Next, we evaluate the compositeness for several dynamically generated
hadronic resonances, which are described not in quark-gluon but in
hadronic degrees of freedom and hence may be hadronic molecules.

Before calculating the compositeness for hadrons, we mention that,
because the compositeness as well as the wave function itself is not a
physical observable, the compositeness is a model dependent
quantity~\cite{Nagahiro:2014mba}.  An exception is the case that the
pole exists very close to the on-shell energies, where we can
model-independently express the compositeness with the scattering
length and effective range~\cite{Weinberg:1965zz, Baru:2003qq,
  Hyodo:2013iga, Kamiya:2015aea}.  Besides, as described above, the
compositeness from the scattering amplitude is uniquely determined
once we fix the model space and interaction.

In the present study we fix the interaction as the separable form for
the orbital angular momentum $L$ ~\cite{Aceti:2012dd}
\begin{equation}
  \langle \bm{q}^{\prime}_{j} | \hat{V} ( E ) | \bm{q}_{k} \rangle
  = ( 2 L + 1 ) q^{\prime \, L} q^{L} 
  P_{L} ( \hat{q}^{\prime} \cdot \hat{q} )
  V_{j k} ( E ) ,
\label{eq:Vsep}
\end{equation}
where $\hat{q}^{( \prime ) } \equiv \bm{q}^{( \prime ) } / | \bm{q}^{(
  \prime ) }|$.  Then, the Lippmann--Schwinger
equation~\eqref{eq:LSint} becomes an algebraic form:
\begin{equation}
  T_{j k} ( E ) = V_{j k} ( E ) + \sum _{l} V_{j l} ( E ) 
  G_{l} ( E ) T_{l k} ( E ) ,
  \quad
  G_{j} ( E ) \equiv \int \frac{d^{3} k}{( 2 \pi )^{3}} 
  \frac{k^{2 L}}{E - \mathcal{E}_{j} ( k )} ,
\end{equation}
where the ultraviolet divergence in the loop function $G_{j} ( E )$ is
regularized in a certain way.  Then, the residue $\gamma _{j}$ becomes
a coupling constant and hence the compositeness
is~\cite{Sekihara:2014kya, Sekihara:2015gvw}
\begin{equation}
  X_{j} = - \gamma _{j}^{2} \left [ \frac{d G_{j}}{d E} 
  \right ]_{E = E_{\rm pole}} .
\end{equation}
In addition, we can define the missing-channel contribution $Z$ as the
rest part for the normalization of the total wave function, {\it
  i.e.}, unity minus the sum of the compositeness:
\begin{equation}
  Z \equiv 1 - \sum _{j} X_{j}
  = - \sum _{j, k} \gamma _{k} \gamma _{j} 
  \left [ G_{j} \frac{d V_{j k}}{d E} G_{k} \right ]_{E = E_{\rm pole}} ,
\end{equation}
where the sum rule for $X_{j}$ and $Z$ was proved in
Ref.~\cite{Sekihara:2010uz}.  We also note that the compositeness and
missing-channel contribution are in general complex for resonances,
which cannot be interpreted as probabilities.  Here, from the complex
$X_{j}$ and $Z$ we introduce quantities which a probabilistic
interpretation is possible:
\begin{equation}
  \tilde{X} \equiv \frac{1 + | X | - | Z |}{2} ,
  \quad 
  \tilde{Z} \equiv \frac{1 + | Z | - | X |}{2} ,
\end{equation}
for the one-channel problem, and
\begin{equation}
  \tilde{X}_{j} \equiv \frac{| X_{j} |}{1 + U} ,
  \quad 
  \tilde{Z} \equiv \frac{| Z |}{1 + U} ,
  \quad
  U \equiv \sum _{j} | X_{j} | + | Z | - 1 ,
\end{equation}
for the problems with more than one channel.  In any case, based on a
similarity between the resonance wave function considered and a wave
function of a stable bound state, with $U \ll 1$ one can safely
interpret $\tilde{X}_{j}$ ($\tilde{Z}$) as the probability of finding
the composite (missing) part.

\def\arraystretch{0.9}

\begin{table}[!t]
  \caption{Compositeness for dynamically generated hadronic resonances:
    scalar mesons. }
  \label{tab:1}
  \centering
  \begin{tabular}{lcccc}
    \hline \hline
    & $f_{0} (500)$~\cite{Sekihara:2014kya}
    & $f_{0} (980)$~\cite{Sekihara:2014kya}
    & $a_{0} (980)$~\cite{Sekihara:2012xp}
    & $K_{0}^{\ast} (800)$~\cite{Sekihara:2014kya}
    \\
    \hline
    $E_{\rm pole}$ [MeV] &
    $443 - 217 i$ &
    $988 - 4 i$ &
    $979 - 53 i$ &
    $750 - 227 i$
    \\
    $X_{\pi \pi}$ &
    $-0.09 + 0.37 i \phantom{-}$ &
    $0.00 - 0.00 i$ &
    --- &
    ---
    \\
    $X_{\pi \eta}$ &
    --- &
    --- &
    $-0.06 + 0.10 i \phantom{-}$ &
    ---
    \\
    $X_{K \bar{K}}$ &
    $-0.01 - 0.00 i \phantom{-}$ &
    $0.87 - 0.04 i$ &
    $0.38 - 0.29i$ &
    ---
    \\
    $X_{\eta \eta}$ &
    $-0.00 + 0.00 i \phantom{-}$ &
    $0.06 + 0.01 i$ &
    --- &
    ---
    \\
    $X_{\pi K}$ &
    --- &
    --- &
    --- &
    $0.32 + 0.36 i$ 
    \\
    $X_{\eta K}$ &
    --- &
    --- &
    --- &
    $-0.01 - 0.00 i \phantom{-}$ 
    \\
    $Z$ &
    $1.09 - 0.37 i$ &
    $0.07 + 0.02 i$ &
    $0.68 + 0.18 i$ &
    $0.70 - 0.36 i$
    \\
    $U$ &
    $0.54$ &
    $0.00$ &
    $0.30$ &
    $0.28$
    \\
    $\tilde{X}_{\pi \pi}$ &
    $0.25$ &
    $0.00$ &
    --- &
    ---
    \\
    $\tilde{X}_{\pi \eta}$ &
    --- &
    --- &
    $0.09$ &
    ---
    \\
    $\tilde{X}_{K \bar{K}}$ &
    $0.01$ &
    $0.87$ &
    $0.37$ &
    ---
    \\
    $\tilde{X}_{\eta \eta} $ &
    $0.00$ &
    $0.06$ &
    --- &
    ---
    \\
    $\tilde{X}_{\pi K}$ &
    --- &
    --- &
    --- &
    $0.38$
    \\
    $\tilde{X}_{\eta K}$ &
    --- &
    --- &
    --- &
    $0.01$
    \\
    $\tilde{Z}$ &
    $0.75$ &
    $0.07$ &
    $0.54$ &
    $0.62$
    \\
    \hline \hline
  \end{tabular}
\end{table}

Now we calculate the compositeness for hadronic resonances.  In this
study we employ chiral perturbation theory for the separable
interaction~\eqref{eq:Vsep}, with which we can dynamically generate
several hadronic resonances.  Here we treat $f_{0} (500)$, $f_{0}
(980)$, $a_{0} (980)$, and $K_{0}^{\ast} (800)$ for mesons and $\Delta
(1232)$, $N (1535)$, $N (1650)$, $\Lambda (1405)$, $\Lambda (1670)$,
and $\Xi (1690)$ for baryons.  In particular, the scattering
amplitudes for the scalar mesons $f_{0} (500)$, $f_{0} (980)$, and
$K_{0}^{\ast} (800)$ are taken from Ref.~\cite{GomezNicola:2001as},
and for $\Lambda (1405)$ from Ref.~\cite{Ikeda:2011pi}.  The results
are listed in Tables~\ref{tab:1}, \ref{tab:2}, and \ref{tab:3}.

First, among the scalar mesons in Table~\ref{tab:1}, the $K \bar{K}$
compositeness for the $f_{0} (980)$ resonance, both $X_{K \bar{K}}$
and $\tilde{X}_{K \bar{K}}$, is close to unity with negligible value
of $U$.  This result suggests that $f_{0} (980)$ in this model is
indeed a $K \bar{K}$ molecular state.  On the other hand, the other
scalar mesons have large contributions from the missing channels, $Z$.
Among them, however, $f_{0} (500)$ has the value of $U$ comparable to
unity, so one cannot clearly interpret the structure of $f_{0} (500)$
from $X$ and $Z$.

%$f_{0} (500)$, $a_{0} (980)$, and $K_{0}^{\ast} (800)$ resonances

\begin{table}[!t]
  \caption{Compositeness for dynamically generated hadronic resonances:
    baryons with $S = 0$. }
  \label{tab:2}
  \centering
  \begin{tabular}{lccc}
    \hline \hline
    & $\Delta (1232)$\footnotemark~\cite{Sekihara:2015gvw}
    & $N (1535)$~\cite{Sekihara:2015gvw}
    & $N (1650)$~\cite{Sekihara:2015gvw}
    \\
    \hline
    $E_{\rm pole}$ [MeV] &
    $1207 - 50 i$ &
    $1496 - 59 i$ &
    $1661 - 70 i$
    \\
    $X_{\pi N}$ &
    $0.87 + 0.35 i$ &
    $-0.02 + 0.03 i \phantom{-}$ &
    $0.00 + 0.04 i$ 
    \\
    $X_{\eta N}$ &
    --- &
    $0.04 + 0.37 i$ &
    $0.00 + 0.01 i$ 
    \\
    $X_{K \Lambda}$ &
    --- &
    $0.14 + 0.00 i$ &
    $0.08 + 0.05 i$ 
    \\
    $X_{K \Sigma}$ &
    --- &
    $0.01 - 0.02 i$ &
    $0.09 - 0.12 i$ 
    \\
    $Z$ &
    $0.13 - 0.35 i$ &
    $0.84 - 0.38 i$ &
    $0.84 + 0.01 i$
    \\
    $U$ &
    $0.31$ &
    $0.48$ &
    $0.13$
    \\
    $\tilde{X}_{\pi N}$ &
    $0.78$ &
    $0.03$ &
    $0.04$
    \\
    $\tilde{X}_{\eta N}$ &
    --- &
    $0.25$ &
    $0.01$
    \\
    $\tilde{X}_{K \Lambda}$ &
    --- &
    $0.09$ &
    $0.08$ 
    \\
    $\tilde{X}_{K \Sigma} $ &
    --- &
    $0.01$ &
    $0.13$
    \\
    $\tilde{Z}$ &
    $0.22$ &
    $0.62$ &
    $0.74$
    \\
    \hline \hline
  \end{tabular}
\end{table}
\footnotetext{We only consider the $\pi N$ channel for the $\Delta
  (1232)$ resonance.}

Next, from the results for the nucleon resonances in
Table~\ref{tab:2}, one can see that the $\Delta (1232)$ resonance has
a large contribution from the $\pi N$ component, as in the previous
work~\cite{Aceti:2014ala}.  The imaginary part of the $\pi N$
compositeness for $\Delta (1232)$ is also non-negligible, but $U$ is
less than one-third, so we may conclude that the $\pi N$ component in
$\Delta (1232)$ is significant.  On the other hand, the results on $N
(1535)$ and $N (1650)$ indicate that the $\pi N$, $\eta N$, $K
\Lambda$, and $K \Sigma$ components are negligible for these
resonances.

\begin{table}[!t]
  \caption{Compositeness for dynamically generated hadronic resonances:
    baryons with $S \le -1$. }
  \label{tab:3}
  \centering
  \begin{tabular}{lcccc}
    \hline \hline
    & $\Lambda (1405)_{\text{higher}}$~\cite{Sekihara:2014kya}
    & $\Lambda (1405)_{\text{lower}}$~\cite{Sekihara:2014kya}
    & $\Lambda (1670)$~\cite{Sekihara:2014kya}
    & $\Xi (1690)$~\cite{Sekihara:2015qqa}
    \\
    \hline
    $E_{\rm pole}$ [MeV] &
    $1424 - 26 i$ &
    $1381 - 81 i$ &
    $1678 - 21 i$ &
    $1684 - 1 i$
    \\
    $X_{\bar{K} N}$ &
    $1.14 + 0.01 i$ &
    $-0.39 - 0.07 i \phantom{-}$ &
    $0.03 + 0.00 i$ &
    ---
    \\
    $X_{\pi \Sigma}$ &
    $-0.19 - 0.22 i \phantom{-}$ &
    $0.66 + 0.52 i$ &
    $0.00 + 0.00 i$ &
    ---
    \\
    $X_{\eta \Lambda}$ &
    $0.13 + 0.02 i$ &
    $-0.04 + 0.01 i \phantom{-}$ &
    $-0.09 + 0.16 i \phantom{-}$ &
    ---
    \\
    $X_{K \Xi}$ &
    $0.00 + 0.00 i$ &
    $-0.00 + 0.00 i \phantom{-}$ &
    $0.53 - 0.10 i$ &
    ---
    \\
    $X_{\bar{K} \Sigma}$ &
    --- &
    --- &
    --- &
    $0.95 - 0.14 i$ 
    \\
    $X_{\bar{K} \Lambda}$ &
    --- &
    --- &
    --- &
    $-0.02 + 0.00 i \phantom{-}$ 
    \\
    $X_{\pi \Xi}$ &
    --- &
    --- &
    --- &
    $0.00 + 0.00 i$
    \\
    $X_{\eta \Xi}$ &
    --- &
    --- &
    --- &
    $0.01 + 0.02 i$ 
    \\
    $Z$ &
    $-0.08 + 0.19 i \phantom{-}$ &
    $0.77 - 0.46 i$ &
    $0.53 - 0.06 i$ &
    $0.06 + 0.11 i$
    \\
    $U$ &
    $0.77$ &
    $1.17$ &
    $0.29$ &
    $0.13$
    \\
    $\tilde{X}_{\bar{K} N}$ &
    $0.64$ &
    $0.18$ &
    $0.02$ &
    ---
    \\
    $\tilde{X}_{\pi \Sigma}$ &
    $0.16$ &
    $0.39$ &
    $0.00$ &
    ---
    \\
    $\tilde{X}_{\eta \Lambda}$ &
    $0.07$ &
    $0.02$ &
    $0.14$ &
    ---
    \\
    $\tilde{X}_{K \Xi} $ &
    $0.00$ &
    $0.00$ &
    $0.42$ &
    ---
    \\
    $\tilde{X}_{\bar{K} \Sigma}$ &
    --- &
    --- &
    --- &
    $0.85$
    \\
    $\tilde{X}_{\bar{K} \Lambda}$ &
    --- &
    --- &
    --- &
    $0.02$
    \\
    $\tilde{X}_{\pi \Xi}$ &
    --- &
    --- &
    --- &
    $0.00$
    \\
    $\tilde{X}_{\eta \Xi}$ &
    --- &
    --- &
    --- &
    $0.02$
    \\
    $\tilde{Z}$ &
    $0.12$ &
    $0.41$ &
    $0.41$ &
    $0.11$
    \\
    \hline \hline
  \end{tabular}
\end{table}

The results for the baryons with $S \le -1$ are shown in
Table~\ref{tab:3}.  An interesting finding is that the higher pole of
$\Lambda (1405)$ has a dominant $\bar{K} N$ compositeness $X_{\bar{K}
  N}$, which is very close to unity.  Although the value of $U$ is
non-negligible compared to unity for the higher $\Lambda (1405)$, this
mainly comes from the imaginary parts and negative real parts of
contributions other than $\bar{K} N$, and $U$ divided by the number of
channels is small: $U / 5 \approx 0.15 \ll 1$.  Therefore, we may
state that the higher $\Lambda (1405)$ is dominated by the $\bar{K} N$
composite component.  The $\Lambda (1670)$ resonance has substantial
$K \Xi$ component in the present model.  Finally, the result on $\Xi
(1690)$ indicates that this resonance is indeed a $\bar{K} \Sigma$
molecular state.

\section{Summary}

We have shown that the two-body wave function of a bound/resonance
state can be evaluated from the scattering amplitudes as an
automatically normalized quantity.  In particular, the compositeness,
{\it i.e.}, the norm of the two-body wave function, is unity for an
energy-independent interaction, while the compositeness deviates from
unity for an energy-dependent interaction, which can be interpreted to
implement a missing-channel contribution.

By evaluating the compositeness for several dynamically generated
resonances with the interaction taken from chiral perturbation theory,
we conclude that $f_{0} (980)$, $\Lambda (1405)$, and $\Xi (1690)$ are
dominated by the $K \bar{K}$, $\bar{K} N$, and $\bar{K} \Sigma$
composite components, respectively, in the present model.

\end{document}